\documentclass{aa}  
\usepackage{graphicx}
\usepackage{natbib}
\bibpunct{(}{)}{;}{a}{}{,}    %to follow the A&A style

\usepackage{txfonts}
\begin{document}
   \title{On the evolution of the resonant planetary system 
          HD~128311}

   \author{Zs. S\'andor
          \inst{1,2}
          \and
          W. Kley\inst{2}
          }

   \offprints{Zs. S\'andor}

   \institute{Department of Astronomy, E\"otv\"os University,
              P\'azm\'any P\'eter s\'et\'any. 1/A, H-1117 Budapest, Hungary\\
              \email{Zs.Sandor@astro.elte.hu}
         \and
            Institut f\"ur Astronomie und Astrophysik, Abt. Computational Physics, Universit\"at T\"ubingen,\\
             Auf der Morgenstelle 10, D-72076 T\"ubingen, Germany\\
	     \email{kley@tat.physik.uni-tuebingen.de}
	     }

   \date{Received ; accepted }

  \abstract
   {A significant number of the known multiple exoplanetary systems are
   containing a pair of giant planets engaged in a low order mean motion
   resonance. Such a resonant condition protects the dynamics of these planets
   resulting in very stable orbits. According to recent studies the capture
   into a resonance is the result of a planetary migration process induced
   by the interaction of the planets with a protoplanetary disk.
   If the migration is slow enough (adiabatic) next to a mean motion
   resonance, the two planets will also be in apsidal corotation.}
   {The recently refined orbital parameters of the system HD~128311
   suggest that the two giant planets are in a 2:1 mean motion
   resonance, however without exhibiting apsidal corotation. Thus the evolution of
   this system can not be described by an adiabatic migration process alone. 
   We present possible evolution scenarios of this system combining migration 
   processes and sudden perturbations.
   }
   {We model migration scenarios through numerical
   integration of the gravitational N-body problem with additional non-conservative forces. 
   Planet-planet scattering has been investigated by N-body simulations.}
   {We show that the present dynamical state of the system HD~128311 may be
   explained by such evolutionary processes.}
   {}


   \maketitle
%
%________________________________________________________________

\section{Introduction}
Among the 19 multi-planet systems found to date about a third are engaged in a low order mean motion
resonance \citep{2005ApJ...632..638V}. The most prominent case is the exact 2:1 resonance of the
two outer planets in GJ~876.  There the orbital elements are very well determined, due to the
short periods of the planets of $\approx$ 30 and 60 days \citep{2005ApJ...622.1182L}.
The formation of resonant configurations
between planets must be due to dissipative processes altering the semi-major axis.
For the system GJ~876 the orbits are in apsidal corotation and both resonant angles
librate with small amplitudes, a condition which can best explained by a
sufficiently slow and long lasting differential migration process induced by the interaction of the
planets with a protoplanetary disk 
\citep{2001A&A...374.1092S, 2002ApJ...567..596L, 2005A&A...437..727K}.
Hence, the occurrence of resonances constitutes a very strong indication of migration
in young planetary systems, in addition to the hot Jupiter cases.

Recent analysis of the planetary system HD~128311 suggests that two giant planets 
are engulfed in a 2:1 mean motion resonance, however without exhibiting apsidal corotation
\citep{2005ApJ...632..638V}.  On the other hand, 
according to \citet{2002ApJ...567..596L, 2006MNRAS.365.1160B}, in the
case of a sufficiently slow migration process the resonant planets should also be in apsidal corotation.
In order to resolve the above discrepancy we construct {\it mixed} evolutionary scenarios of migrating planetary 
systems incorporating migration and other additional perturbative effects.
For our investigation we use N-body numerical 
integrations containing also non-conservative forces \citep{2002ApJ...567..596L},
which have been tested on full hydrodynamic evolutions of embedded planets \citep{2004A&A...414..735K}.
In this letter we report our findings in modeling the behavior of the resonant exoplanetary system HD~128311. 

\section{Orbital solution and its stability}
It is mentioned by \citet{2005ApJ...632..638V} that the best fit to the
radial velocity curve of HD~128311,
based only on the use of Keplerian ellipses, leads to orbital data resulting
in unstable behavior. Therefore they present also an alternate fit for HD~128311
calculated by using three-body gravitational interactions. The orbital data
determined by this alternate fit (listed in Table 1) result in stable orbits,
where the two giants are in a protecting 2:1 mean motion resonance (MMR).
We perform three-body numerical integrations using these initial conditions of 
\citet{2005ApJ...632..638V}, as listed in Table 1. 
Only the resonant angle $\Theta_1 = 2\lambda_2 - \lambda_1 - \varpi_1$ 
librates around $0^{\circ}$ with an amplitude of
$\sim 60^{\circ}$ (Fig.~\ref{Fig1}, top), while both 
$\Theta_2 = 2\lambda_2 - \lambda_1 - \varpi_2 $ and 
$\Delta \varpi = \varpi_2 - \varpi_1$ circulate.
Here $\lambda_j$ are mean longitudes and $\varpi_j$ are
longitudes of periapse, both numbered from the inside out. 
This implies that the systems, although engaged in a MMR, is not in apsidal corotation.
Moreover, the eccentricities show also large oscillations (Fig.~\ref{Fig1}, bottom).

As described by \citet{2005ApJ...632..638V}, due to the relatively high stellar jitter,
there exist a large variety of dynamically distinct stable orbital solutions resulting
in the current radial velocity curve. The authors present the
results obtained by a self-consistent two-planet model (taking into account the mutual
gravitational interactions) and a Monte-Carlo procedure. As final outcome,
they have found that only very few samples exhibit stable behavior without resonance.
The remaining stable solutions are in 2:1 MMR, in the majority of the cases
($\sim 60\%$) only $\Theta_1$ librates while $\Theta_2$ circulates, in the remaining
cases $\Theta_2$ librates as well. Hence, we assume that their derived
parameters in Table~1 are an accurate representation of the dynamics of the system.

\begin{table}
\caption{Orbital data of HD~128311 as provided
by \citet{2005ApJ...632..638V}. $M$ denotes the mean anomaly.}
\label{table:1}      
\centering                         
\begin{tabular}{c c c c c c}       
\hline\hline              
planet & mass [M$_J$]& $a$ [AU] & $e$ & $M$ [deg] & $\varpi$ [deg] \\   
\hline                      
   b  & 1.56 & 1.109 & 0.38 & 257.6 & 80.1 \\     
   c  & 3.08 & 1.735 & 0.21 & 166.0 & 21.6  \\
\hline                                   
\end{tabular}
\end{table}
\section{Evolution through a migration process}
\subsection{Adiabatic migration}
\label{sec:adiabat}
As shown in the previous section, the planets in the system HD~128311
are presently engaged in a stable
2:1 MMR. Although the actual orbital parameters do not
exhibit apsidal corotation, the existence of the resonance suggests
that the system has in the past gone through a migration process.

The migration of a single planet can be characterized by the migration rate $\dot a/a$ and
the eccentricity damping rate $\dot e/e$. Here, we use the corresponding
e-folding times for the semi-major axes and eccentricities: $\tau_a$ and $\tau_e$,
respectively. The relation between the damping rates and e-folding times is
$\dot a/a = - 1/\tau_a$, and similarly for the eccentricities.
Investigating the system GJ~876, \citet{2002ApJ...567..596L} have found that for a
sufficiently slow migration the final state of the system depends only on the
ratio $K$ of the e-folding times $K = \tau_a/\tau_e$.

From hydrodynamical calculations we know that the order of magnitude of $K$ is close
to unity and it reflects the physical properties (i.e. mass and viscosity)
of the protoplanetary disk \citep{2004A&A...414..735K}. 
If a system is a subject to such an adiabatic migration, a given value of
$K$ results in unique values of the final eccentricities \citep{2002ApJ...567..596L}.
However, as shown above, in the case of HD~128311 the eccentricities are varying with quite large
amplitudes.

We have performed a set of numerical integrations of the
general co-planar three-body problem adding non-conservative drag forces,
varying the value of $K$.

\begin{figure}
   \centering
   \includegraphics[width=8cm]{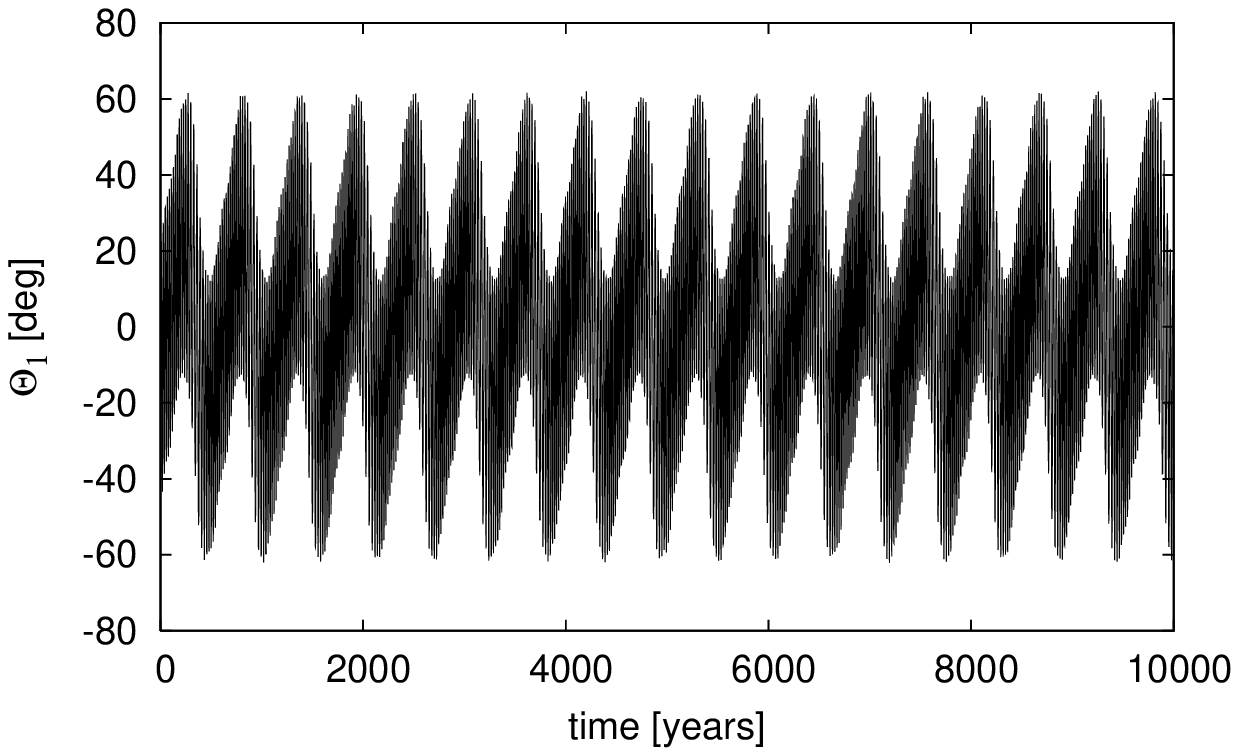}
   \includegraphics[width=8cm]{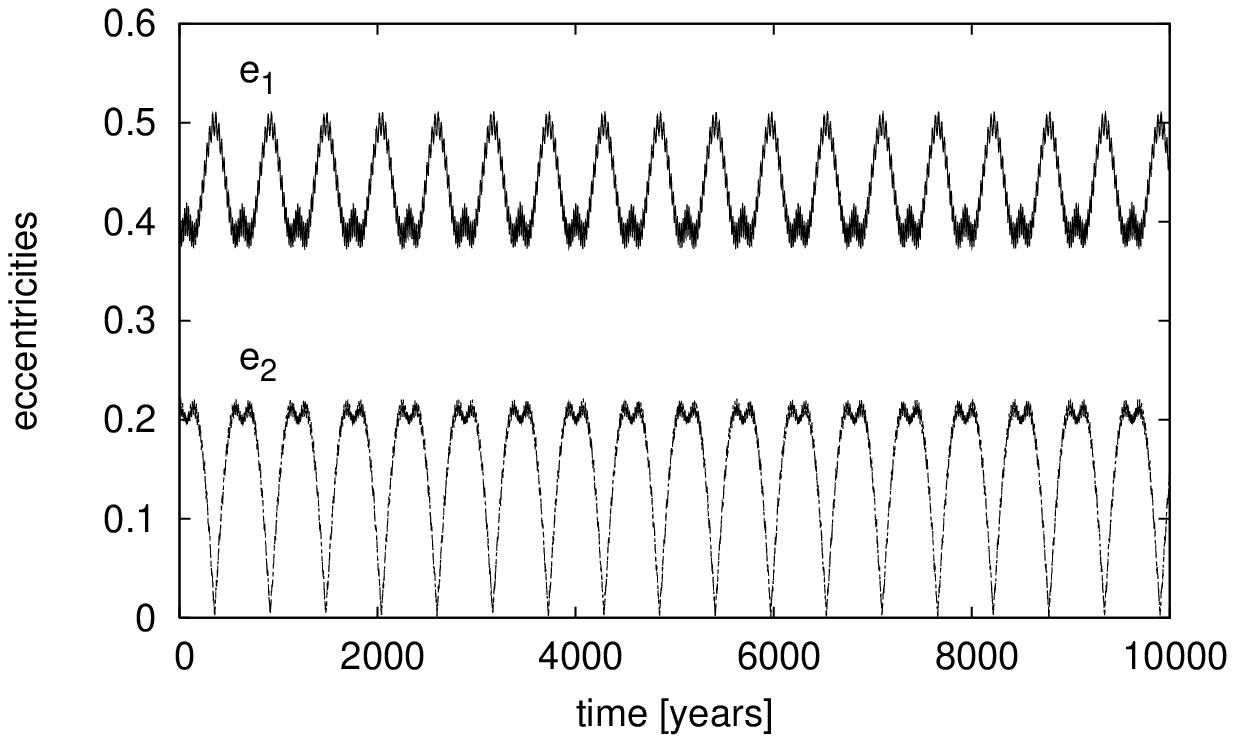}
      \caption{{\bf Top}: Time evolution of the resonant angle $\Theta_1$ obtained by
       numerical integration using initial conditions of Table 1.
      {\bf Bottom}: The corresponding evolution of the eccentricities.}
         \label{Fig1}
   \end{figure}

\begin{figure}
   \centering
   \includegraphics[width=8cm]{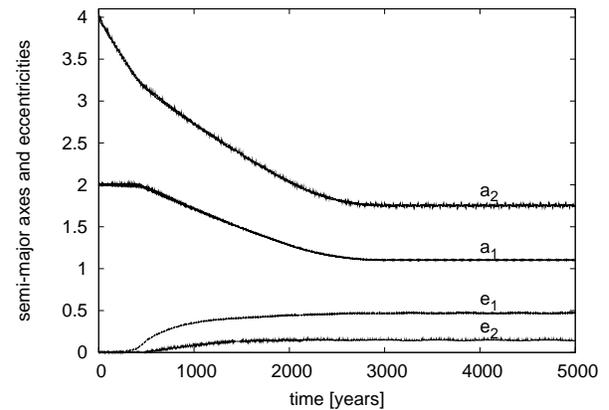}
      \caption{The behavior of the semi-major axes and eccentricities during an adiabatic
               migration with $\tau_a=2\times10^3$ years and $K=5$. The migration is stopped
	        between $2\times10^3$ and $3\times10^3$ years by applying a linear reduction.}
         \label{fig:adiabat}
   \end{figure}
   
In these simulations the planets move originally in quasi-circular
orbits and we start them from $a_1=4$AU and $a_2=2$AU. Assuming that only
the outer planet is forced to migrate inward, we implement
a dissipative force which results in a migration characterized by the
e-folding time $\tau_a=2\times10^3$ years in the semi-major axis, and
$\tau_e=4\times10^2$ years in the eccentricity of the outer planet 
(corresponding to $K=5$). 
In order to obtain the present values of the semi-major
axes we slow down the migration beginning from $t_1=2\times10^3$years and decreasing
it linearly to zero until $t_2=3\times10^3$years. In this way we model the smooth
dispersal of the protoplanetary disk. After this migration process, the system
is locked into a deep 2:1 resonance, the resonant angles $\Theta_{1,2}$, and 
$\Delta \varpi$ librate around $0^{\circ}$ with small amplitudes.
The eccentricities $e_1$ and $e_2$ are almost constant
(showing only a small amplitude oscillations),
where the index 1 refers to the inner and the index 2 to the outer planet.
For the chosen $K=5$ we find $e_1=0.46$ and $e_2=0.15$, see Fig.~\ref{fig:adiabat}.
Smaller/larger values of $K$ yield always systems deep in resonance
and result in larger/smaller $e_{1,2}$, contradicting Fig.~\ref{Fig1}.

{\it Clearly, the present behavior of HD~128311 is not the result of such an
adiabatic migration process alone.} In the following, we shall present two additional
mechanisms, which may be responsible for the large oscillations of the eccentricities and
breaking the apsidal corotation. 
\subsection{Sudden stop of the migration of the outer planet}
Recent {\it Spitzer} observations of young stars confirm that the inner part of the protoplanetary disk may
contain only very little mass \citet{2005ApJ...621..461D, 2005ApJ...630L.185C}, possibly 
due to photo-evaporation induced by the central star. Thus, upon approaching the inner rim of such a disk, 
the inward migration of a planet can be terminated rapidly.  

\begin{figure}
   \centering
   \includegraphics[width=8cm]{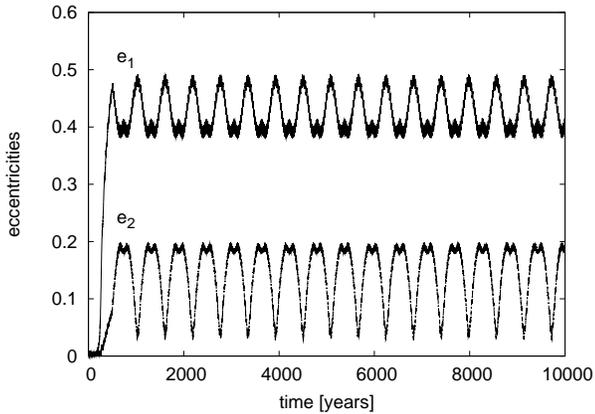}
      \caption{The behavior of the eccentricities obtained by a sudden stop of the migration of the outer planet.}
         \label{fig:sudden}
   \end{figure}

In order to model this type of scenario, we perform additional simulations where the migration of the 
outer planet has been stopped abruptly reaching the actual value of its semi-major axis ($a_2=1.73$AU). 
We assume that the inner planet orbits in the empty region of the disk at $a_1=1.5$AU, and we start
the outer planet from $a_2=4$AU forcing to migrate inward very fast ($\tau_a = 500$ years). We find 
that the present behavior of the eccentricities can be obtained by using $K=10$. 
%%  which may be realistic for thick disks. 
The sudden stop of the migration results in a behavior of the eccentricities shown in 
Fig.~\ref{fig:sudden} which is very similar to the observed case (Fig.~\ref{Fig1}).
We note that after the above sudden stop of the migration the planets remain in apsidal corotation but with 
increased libration amplitudes of the resonant angles.  
%Figure 4
%
\subsection{Planet-planet scattering event}
\begin{figure}
   \centering
   \includegraphics[width=8cm]{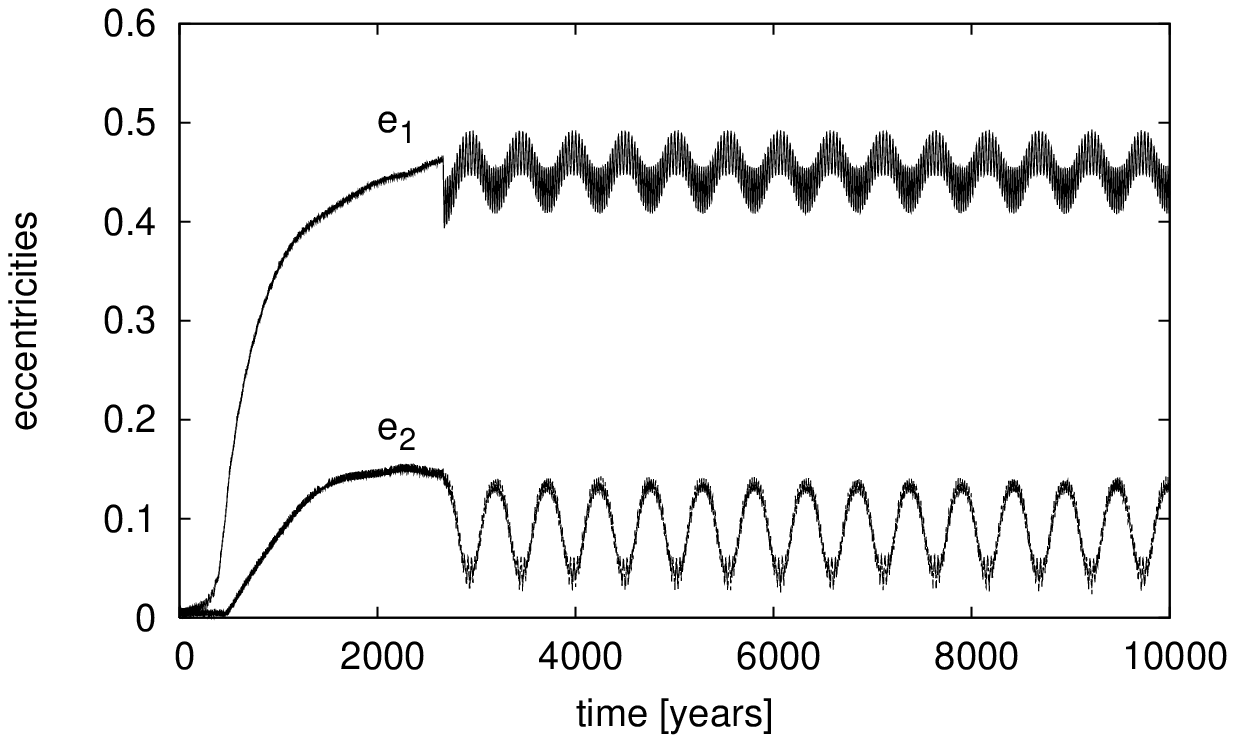}
   \includegraphics[width=8cm]{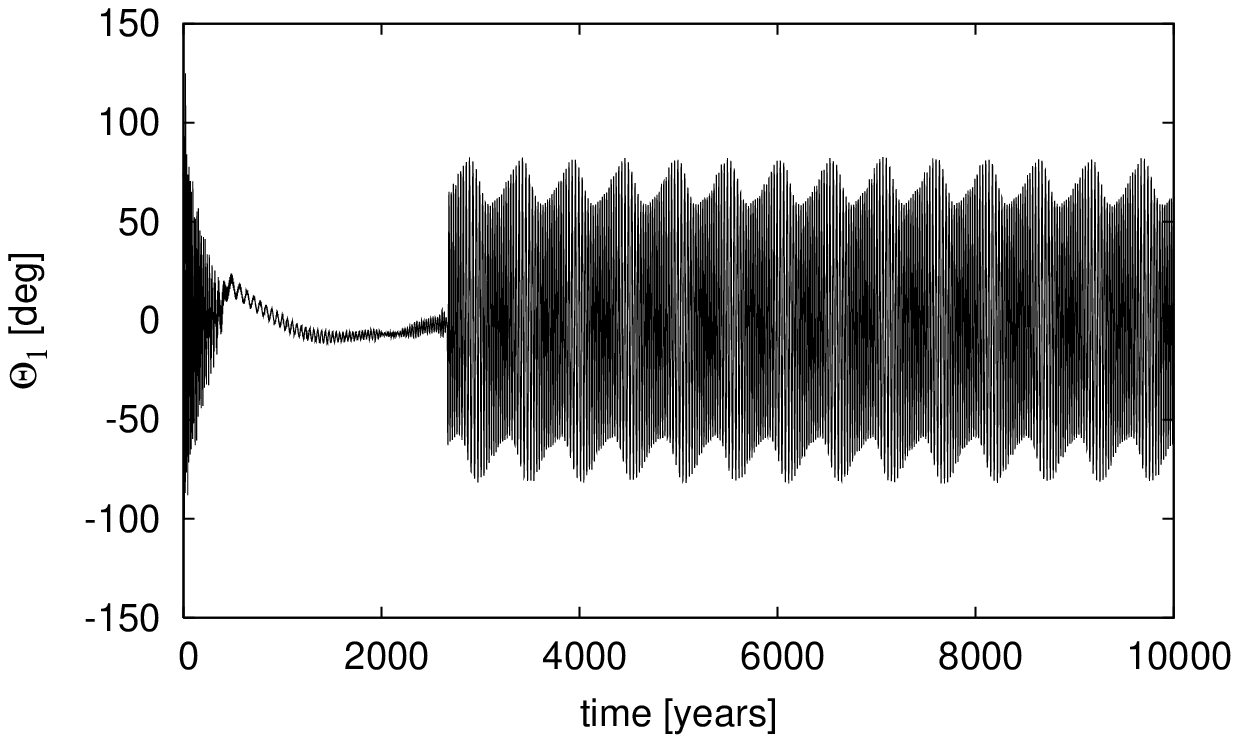}
   \includegraphics[width=8cm]{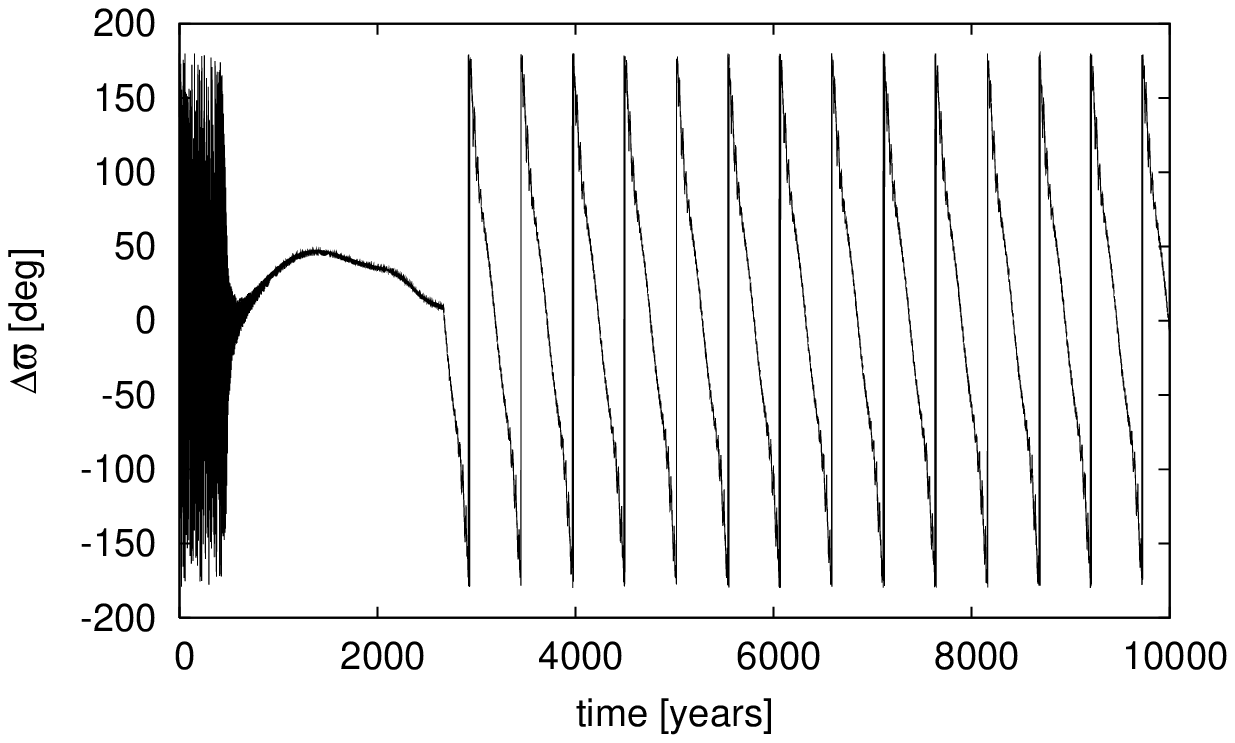}
      \caption{
    Time evolution of the eccentricities ({\bf Top}), 
    resonant angles $\Theta_1$ ({\bf Middle}) 
     and $\Delta\varpi$ ({\bf Bottom}) of the 
   inward migrating giant planets,
   before and after a scattering event with an inner low mass planet.
	       }
         \label{fig:inner1}
\end{figure}  
The behavior of the eccentricities of the giant planets of the system HD~128311
is very similar to that observed in the system around $\upsilon$ Andromedae.
\citet{2005Natur.434..873F} proposed that such a behavior in $\upsilon$ And is most likely the result
of a planet-planet scattering event. Investigating the resonant system HD~73526, 
\citet{tinney-2006-} also suggested that the lack of the apsidal corotation can be the result
of a dynamical scattering event. Therefore we investigate whether the present
behavior of HD~128311 can be modeled by such an effect. 
We present two cases.\\[0.1cm]
{\it Additional inner planet} \\[0.1cm]
In this case we assume that an additional 
small mass planet is orbiting close to the central star in a quasi circular orbit. 
%% These assumption is reasonable, since if the mass and the orbital eccentricity
%% of this planet would be larger, it would already be detected by the radial
%% velocity measurements.
When the outer giants are far enough from the star the orbit of the small mass
planet is stable. However, as the giants migrate inward approaching their
present positions they perturb the orbit of the small planet making its motion
chaotic, which in long term may result in an increase of its eccentricity.
Finally, due to its high eccentricity, the small planet can suffer a close
encounter with one of the giant planets. After the close encounter the small
planet may be ejected from the system or pushed into an orbit with very large
semi-major axis. Depending on the initial position of the small planet, it can
also be captured into a mean-motion resonance with the inner giant, and forced
to migrate inward. During the migration its eccentricity will increase in a
much shorter timescale than by a ``pure" chaotic evolution. Thus, in this case a
close encounter with the inner giant is very likely as well.

\begin{figure}
   \centering
   \includegraphics[width=8cm]{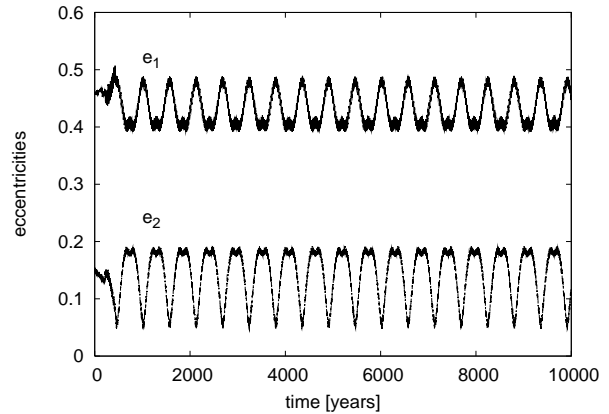}
  \caption{The behavior of the eccentricities before and after a scattering with a small mass 
     planet migrating inward from outside starting from $a=2.6$AU.}
       \label{fig:outer1}
   \end{figure}

In Fig.~\ref{fig:inner1} we show the behavior of the eccentricities and the resonant angles
$\Theta_1$ and $\Delta\varpi$  when the giant planets are migrating inwards
with initial conditions given in Sect.~\ref{sec:adiabat}.
Starting the small planet ($m=0.03M_{\rm Jup}$) from a nearly circular orbit with $a=0.5$AU,
it is captured into a resonance with the inner giant. During the migration the small
planet's eccentricity increases, and finally the small planet crosses the orbits of
giant planets and suffers a close encounter with them. After the encounter
at $t\approx 2500$yrs, the apsidal
corotation of the giant planets breaks (i.e. $\Theta_2$ and $\Delta \varpi$ circulate), however the giant planets 
remain in the 2:1 resonance, since $\Theta_1$ still librates around $0^\circ$,
see Fig.~\ref{fig:inner1}.\\[0.1cm]
{\it Additional outer planet} \\[0.1cm]   
In this second case we assume that the third planet ($m=0.03M_{\rm Jup}$) originates from the outer
region, and that it approaches the giant planets through inward migration where we
assume that the adiabatic migration of the two giant planets is already finished,
as displayed in Fig.~\ref{fig:adiabat}.
They orbit for instance in a gas-free environment while the small planet is still embedded
in the outer protoplanetary disk. 
In Fig.~\ref{fig:outer1} we display the variation of the eccentricities of the
giants after the scattering event.
The small planet is started from $a=2.6$AU and migrates
inward with an e-folding time $\tau_a=2\times10^3$ years.
After scattering it is pushed into a very distant orbit $a\sim300$AU.
We note that in this case the giant planets remain in apsidal corotation,
however with a substantial increase in $\Theta_2$ and $\Delta\varpi$. 
In additional simulations we find that apsidal resonance is always preserved,
however we cannot exclude the possibility of breaking the apsidal corotation
entirely.
\section{Conclusions}
In this paper we investigate evolutionary scenarios, which may have led to the observed
behavior of the resonant system HD~128311, where the two planets are engaged in a 2:1 mean motion 
resonance, but not in apsidal corotation. We assume that the two giant planets have been formed far 
from the central star and migrated inwards, due to gravitational interaction with the protoplanetary disk.
During this differential migration, they have been locked into the 2:1 mean motion resonance. 
At the end of their migration we imagine that the system suffers a sudden perturbation. 

As a first case we study the sudden stop of the migration, which may possibly be induced 
by an inner rim of the disk and an empty region inside of it, as indicated by some observations of 
young stars.

In the second scenario the sudden perturbation is caused by a planet-planet scattering event,
similar to that suggested by \citet{2005Natur.434..873F} in the case of $\upsilon$ And.
We analyze an encounter with a small ($\sim 10 M_{\oplus}$) planet,
approaching the two massive planets either from outside or inside, which is
thrown to a large $a$ orbit or directly ejected after the scattering event.

We find that the sudden perturbation caused by an encounter with an {\it inner}
small planet is clearly able to break the apsidal resonance of the two planets.
But also in the other cases the orbital behavior of the giants is very similar
to the most recent orbital data of HD~128311 \citep{2005ApJ...632..638V}.

The system HD~128311 constitutes another example which demonstrates the
important interplay of migration
and scattering processes in shaping the dynamics of exoplanetary systems.

\begin{acknowledgements}
      The authors acknowledge the supports of the Hungarian Scientific Research Fund
      (OTKA) under the grant D048424 and the German Science Foundation (DFG)
      under grant 436 UNG 17/5/05. 
      WK would like to thank Greg Laughlin for the many fruitful discussions.
\end{acknowledgements}

\bibliographystyle{aa}
%%\begin{thebibliography}
\bibliography{kley8}

\end{document}